\documentclass[11pt, a4paper]{article}
\usepackage{authblk}

\usepackage{hyperref}
\usepackage{hypcap}
\usepackage{graphicx}
\usepackage{tikz}
\usepackage{caption}
\usepackage{enumerate}
\usepackage{arydshln}
\usepackage{amssymb}
\usepackage{bbm}
\usepackage{amsmath}
\usepackage{amsfonts}
\usepackage{physics}
\usepackage{nameref}
\usepackage{subfig}
\usepackage{soul}
\usepackage{ulem}
\usepackage{booktabs}
\definecolor{aliceblue}{rgb}{0.94, 0.97, 1.0}
\definecolor{yellow}{rgb}{1, 1, 0.5}
\definecolor{sienna}{rgb}{1, 1, 1}
\definecolor{orange}{rgb}{1, 1, 1}
\usepackage{xcolor,colortbl}
\usepackage{algorithm}
\usepackage{algpseudocode}
\newcolumntype{B}{>{\columncolor{aliceblue}}r}

\usepackage{marginnote}
\setulcolor{red}

\providecommand{\keywords}[1]{
  \small\textbf{\textit{Keywords }} #1
}

\title{Electricity Load and Peak Forecasting: Feature Engineering, Probabilistic LightGBM and Temporal Hierarchies}

\author[a]{Nicolò Rubattu}
\author[a]{Gabriele Maroni}
\author[a]{Giorgio Corani}
\affil[a]{
Dalle Molle Institute for Artificial Intelligence (IDSIA), 
USI-SUPSI,
CH-6962, Lugano, Switzerland
}
\affil[ ]{\small \texttt {\{nicolo.rubattu, gabriele.maroni, giorgio.corani\}@idsia.ch}}

\date{}

\begin{document}
\maketitle
\begin{abstract}
%
%
We describe our experience in developing a predictive model that placed high position in the BigDeal Challenge 2022, an energy competition of load and peak forecasting.
We present a novel procedure for feature engineering and feature selection, based on cluster permutation of temperatures and calendar variables.
We adopted gradient boosting of trees and we enhance its capabilities with trend modelling and distributional forecasts.
We also include an approach to forecast combination known as temporal hierarchies, which further improves the accuracy.

\vspace{3pt}
    \keywords{Load Forecasting\quad Feature engineering\quad Gradient Boosting\quad  Hierarchical Forecasting\quad Temporal reconciliation\quad BigDEAL Challenge 2022}
\end{abstract}

\section*{Introduction}
\label{section:introduction}
\textit{Load forecasting} is the problem of predicting the future profile of power demand, while
\textit{peak forecasting} is the problem of predicting the maximum (e.g. daily) value of  demand and the  time of its occurrence. 
Peak forecasting is important because often decision making is based on the forecast of the peak rather then on the forecast of the entire load profile. 

In this work we present an approach that successfully competed in the BIGDeal Challenge 2022 \cite{BigDEALChallInit}, which
was about energy load and peak forecasting. 
The competition was held in October-December 2022;
121 contestants  took part,
divided in 78 teams.   The forecasts were assessed using different indicators and the competition
was split into  a qualifying match and a final match.
We achieved the $3^{rd}$ position in the qualifying match, gaining access to the final match, where we ended $6^{th}$ \cite{BigDEALChallFinal}.

For the qualifying match we used
Gradient Boosting (GB) of trees, coupled with an original method
of feature engineering and  feature selection.
For the final match, we developed a more sophisticated approach.
In particular, we adopted a recent  probabilistic version of   
LightGBM \cite{Marz2022};
we used the  DART algorithm \cite{vinayak2015dart} for regularization,
and we also used temporal hierarchies \cite{athanasopoulos2017forecasting} in order
to improve the forecasts by combining  predictions at different temporal scales.
Even though the competition only scored the point forecasts, our approach is probabilistic, and thus quantifies the uncertainty of the forecasts. 
This is indeed needed to  support decision making.

The paper is organized as follows. 
Next an outlook of long-term load forecasting and our motivations are given. We introduce Gradient Boosting (GB) of trees and probabilistic extensions in Sec. \nameref{sec:gb}. 
We present our approach for feature engineering for load forecasting in Sec. \nameref{sec:feature_eng}, and feature selection in Sec. \nameref{sec:feature_selection}.
Temporal hierarchies are presented in Sec. \nameref{sec:temphier}.
In Sec. \nameref{sec:casestudy} we detailed review our pipeline with technical insights, and competition results. 
We end this work with a critical \nameref{sec:conclusion}.

\section*{Methodology}
\subsection*{Long-term Load Forecasting}
Load forecasting is the problem of predicting
the electricity demand of the next $H$ time steps, denoted by
$[y_{T+1},\,\ldots\,,y_{T+H}]$.
When  the order of magnitude of $H$ is a few hundreds or more, 
we talk about \textit{long-term} forecasting.  For instance, forecasting a year-ahead at hourly scale implies producing 24 $\times$ 365 = 8760 forecasts.
Classical forecasting strategies
\cite{bontempi2013machine} condition the forecast on 
the last observations of the time series.
However, this is not viable for long-term forecasting, since in this case $y_{T+H}$ is independent from $y_T$.
Long-term forecasting is  better addressed as a 
regression problem, adopting  a rich set of explanatory variables (\textit{features})
\cite{hyndman2009density, fan2011short, charlton2014refined, wang2016electric} 
regarding calendar effect (the day of the week, hour of the day, etc),  holidays, temperature, etc.
This approach allows adopting regression methods 
such as  Gradient Boosting (GB) of trees \cite{friedman2002stochastic}, which is indeed
successful in
long-term energy forecasting  \cite{taieb2014gradient}.




\subsection*{Gradient Boosting and Distributional forecasts}
\label{sec:gb}
In fact, GB achieved top positions in the Global Energy Forecasting Competitions (GEFCom) of 2012, 2014, 2017  \cite{hong2014global, hong2016probabilisticgef, hong2019global}, in the M5 forecasting competition \cite{makridakis2022m5, januschowski2022forecasting} and in 
competitions on tabular data \cite{carlens2022,shwartz2022tabular}.
The most popular implementations are XGBoost \cite{chen2016xgboost}, LightGBM \cite{ke2017lightgbm} and CatBoost \cite{dorogush2018catboost}.
They have comparable accuracy, but LightGBM is generally faster and scales better on large data sets. 

GB can be trained with different loss function besides the traditional least squares.
For instance, GB trained to perform quantile regression won the GEFCom2014 probabilistic competition \cite{gaillard2016additive}.
Yet, even quantile regression  only returns a   point forecasts without a predictive distribution.
It is possible to train different GB models, one for each desired quantile; but if the predicted quantiles cross, the  predictive  distribution is invalid \cite{Meinshausen2006, JMLR:v23:21-0247}.
The recent 
versions of probabilistic  GB of trees constitute a sounder approach \cite{duan2020ngboost, marz2019xgboostlss, Marz2022} to
 probabilistic forecasting. 
In this paper we adopt the model of März et al. \cite{Marz2022}, which returns a multivariate output containing the moments of the predictive distribution.

A successful implementation of GB requires anyway paying attention to some possible issues.
For instance, GB is is generally unable to model a long-term trend.
If the time series is trendy,  it is  recommended  to detrend it,  fit the GB model, and  then add the predicted trend to the GB forecast  \cite{ziel2019quantile, lu2020short, wang2021short}.
Another pre-processing step which is sometimes helpful is a logarithmic transformation which stabilizes the variance of the target time series \cite{smyl2019machine}. 
Moreover, GB is subject to overfitting.
The DART algorithm \cite{vinayak2015dart} solves the problem introducing Dropout regularization, analogously to Neural Networks.

\subsection*{Feature Engineering}
\label{sec:feature_eng}
The exogenous variables that are frequently used in load forecasting are related to calendar and temperatures.
\paragraph{Calendar-based features}
Calendar variables allow to capture the seasonal patterns.
They are commonly  modeled by categorical variables, using one category for each  value of the feature.
For example,
the day of the week is represented
by a categorical variable with seven levels.
Holidays are represented by  a binary variable with two levels: 1 for holiday and 0 for non-holiday.

\paragraph{Lagged and rolling  temperatures}
Temperature  impacts on energy consumption, by driving the 
use of heating, ventilation, and air conditioning (HVAC) systems.
However, there is generally a delay between the change in temperature and the change in energy consumption. 
We thus consider the lagged hourly temperatures:
\begin{equation}
    T(t-h), \qquad h = 1,2,\,\dots\,,L
    \label{eq:lagged_temperatures}
\end{equation}
where $L$ is the maximum  lag, and moving/rolling temperature's statistics: 
\begin{equation}
    T_f^w(t) = f(T(t-1),\,\dots\,, T(t-w))
    \label{eq:rolling_lagged_temperatures}
\end{equation}
where $f(\cdot)$ is some statistical function and $w$ indicates the width of the window of past values of hourly temperatures considered. For example, the moving average of the last 24 hours of temperature values is  $T_{avg}^{24}(t) = \frac{1}{24}\sum_{h=1}^{24}{T(t-h)}$.

\paragraph{Aggregated indicators of temperature}
Aggregated features can capture longer-term effect of temperature on energy load, 
%
smoothing out noise in the data. 
They
can be expressed as $\tilde{T}_{f}^g(t)$ where $g$ is the aggregation period and $f(\cdot)$ is the aggregation function. 
This features includes, for example, the daily maximum and minimum values of the temperature or the monthly standard deviation of the temperature.


In this paper we propose additional aggregation functions (Tab. \ref{tab:signal}) borrowed from signal processing \cite{ericsti2010wavelet, yu2011bearing}, which to the  best of our knowledge have been not yet used  in 
energy forecasting.
The signal processing features should be computed on the  time series of temperature, rescaled to have zero mean in the selected aggregation period.
They provide insights about the variability and shape of the temperature profile within the aggregation period. 
For example, the crest factor  measures the peak-to-average ratio of a signal.
A high daily crest factor corresponds to large variations of  temperature during the day, which generally increase energy demand.
On the other hand, a low daily crest factor corresponds to stable temperatures during the day, which generally decreases energy demand.

\begin{table}[!ht]
\centering
\caption{Signal Processing features for load forecasting.\label{tab:signal}}
\vspace{6pt}
\begin{tabular}{cc}
\midrule
RMS & $x_{RMS} = \sqrt{\frac{1}{N}\sum_{i=1}^N x_i^2}$ \\[10pt]
Peak value & $x_p = \max(\left|x_i\right|)$ \\[10pt]
Crest factor & $x_{crest} = \frac{x_p}{x_{RMS}}$ \\[10pt]
Impulse factor & $x_{if} = \frac{x_p}{\frac{1}{N}\sum_{i=1}^N{\left|x_i\right|}}$ \\[10pt]
Margin factor & $x_{mf} = \frac{x_p}{\left(\sum_{i=1}^N\left|x_i\right|^{1/2}\right)^2}$ \\[10pt]
Shape factor & $x_{sf} = \frac{x_{RMS}}{\frac{1}{N}\sum_{i=1}^N{\left|x_i\right|}}$ \\[10pt]
Peak to peak value & $x_{pp} = \max(x_i) - \min(x_i)$ \\[4pt]
\bottomrule
\end{tabular}
\end{table}

\subsection*{Feature Selection}
\label{sec:feature_selection}
Feature engineering generally generates a  large
set of features, many of which are eventually redundant.
Hence, feature selection is needed to: shorten the training times \cite{li2017feature}, reduce  overfitting \cite{butcher2020feature}, improve the model  interpretability \cite{james2013introduction}.

We perform feature selection based on hierarchical clustering 
and pairwise correlation of the features.
The core block of our approach is
Permutation Feature Importance (PFI),
which measures the drop in performance when a feature is randomly shuffled. It has foundations in \cite{breiman2001random, molnar2020interpretable}.
The size of the drop of performance shows how much the model relies on that feature for predictions.
PFI is appealing since:
\begin{itemize}
    \item[-] it can be applied to any model;
    \item[-] it is easy to implement (Algorithm \ref{alg:PFI});
    \item[-] it can measure feature importance  on the  metric of the competition;
    \item[-] it can be computed out-of-sample.
\end{itemize}

\begin{algorithm}
\caption{Permutation Feature Importance}\label{alg:PFI}
\begin{algorithmic}
\Require A trained model and recorded score $s$ on an evaluation dataset.
\For{feature $x_j, j=1,\dots,d$}
\For{each repetition $k, k=1,\dots,K$}
    \State Randomly shuffle column $j$ of the original evaluation set.
    \State Compute new score $s_{k,j}$ of the model on the perturbed set.
    \EndFor
\State Compute importance of feature $x_j$ as:
$$
I_j = s - \frac{1}{K} \sum_{k=1}^K{s_{k,j}}
$$
\EndFor
\end{algorithmic}
\end{algorithm}

However, shuffling a single features can produce unrealistic samples if features are dependent. Furthermore, correlated features share importance, therefore their relevance may be underestimated (substitution effect) \cite{gregorutti2017correlation, molnar2020interpretable}.

\paragraph{Clustered Permutation Feature Importance}
To overcome such issues we propose a novel approach, which we call Clustered Permutation Feature Importance (CPFI). The method works as follows. At first, groups of highly correlated features are identified by applying hierarchical clustering on the correlation matrix of the features. To form the correlation matrix, a measure of dependence between each feature pair is computed using a correlation index, Pearson's or Spearman's for instance. Then, we jointly shuffle all the variables of the same cluster and we compute the subsequent performance drop. The more orthogonal is the information contained in different clusters, the more reliable is the estimate of importance.

Finally, we drop non-informative feature clusters from the model. On the contrary, we keep in the model the features of the relevant clusters. Also, only one or few features can be selected from each group based on some measure of explanatory with respect to the target, or some expert advice.
We propose a criterion to discriminate informative from non-informative cluster of features in Sec. \nameref{sec:casestudy}.
Similar approaches have already been proposed in finance \cite{de2018advances} and biochemistry \cite{maroni2022informed}.

\subsection*{Temporal hierarchies}
\label{sec:temphier}
\begin{figure}[ht!]
	\centering
 \vspace{6pt}
    \begin{tikzpicture}
        \node[shape=circle,draw=black] (4h1) at (2.25,2.2) {$\hat{h}_{1234}$};
        \node[shape=circle,draw=black] (2h1) at (0.75,1.1) {$\hat{h}_{34}$};
        \node[shape=circle,draw=black] (2h2) at (3.75,1.1) {$\hat{h}_{12}$};
        \node[shape=circle,draw=black, fill=aliceblue] (h1) at (0,0) {$\hat{h}_1$};
        \node[shape=circle,draw=black, fill=aliceblue] (h2) at (1.5,0) {$\hat{h}_2$};
        \node[shape=circle,draw=black, fill=aliceblue] (h3) at (3,0) {$\hat{h}_3$};
        \node[shape=circle,draw=black, fill=aliceblue] (h4) at (4.5,0) {$\hat{h}_4$};

        \path [-] (4h1) edge node[left] {} (2h1);
        \path [-] (4h1) edge node[left] {} (2h2);
        \path [-] (2h1) edge node[left] {} (h1);
        \path [-] (2h1) edge node[left] {} (h2);
        \path [-] (2h2) edge node[left] {} (h3);
        \path [-] (2h2) edge node[left] {} (h4);
	\end{tikzpicture}
    \captionsetup{type=figure}
	\caption{Temporal hierarchy for forecasting 4-hours ahead, using hourly forecasts (bottom level), 2-hourly forecasts and 4-hourly  forecasts.}
	\label{fig:hierExample}
\end{figure}
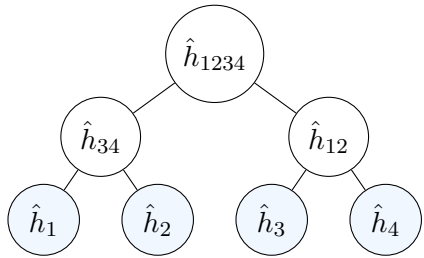

As a further tool to improve the forecasting accuracy, we consider temporal hierarchies \cite{athanasopoulos2017forecasting}.
For instance, assume that we want to  generate forecasts at the hourly scale (referred to as the \textit{bottom level}).
A temporal hierarchy creates forecasts also at coarser temporal scales (e.g., 2-hourly and 4-hourly) and then combines 
information from the forecasts generated at the different temporal scales. This process generally improves the  forecasting  accuracy at all levels \cite{athanasopoulos2017forecasting, KOURENTZES-elucidate}. 
A temporal hierarchy works as follows.
First, forecasts are independently created 
at the different temporal scales (\textit{base forecasts}).
For instance, Fig. \ref{fig:hierExample} shows a 
temporal hierarchy aimed at forecasting 4-hours ahead. It contains 4 forecasts computed at hourly frequency ($\hat{h}_1, ...,  \hat{h}_4$, bottom level);
two  forecasts computed at 2-hour frequency ($\hat{h}_{12}, ...,  \hat{h}_{34}$, intermediate level); one  forecast computed at 4-hour frequency ($\hat{h}_{1234}$, top level).

Generally the base forecasts do not  
sum up correctly and they are referred to as \textit{incoherent}.
For instance:
$\hat{h}_{12} \neq \hat{h}_1 + \hat{h}_2$,
$\hat{h}_{34} \neq \hat{h}_3 + \hat{h}_4$, etc.
\textit{Reconciliation} \cite{Wickramasuriya2019} is the process of adjusting the base forecast so that they become  \textit{coherent}, i.e., they
sum up correctly. 
The reconciled forecasts are denoted with a tilde and thus in the example of Fig. \ref{fig:hierExample} after reconciliation we have: $\tilde{h}_{12} = \tilde{h}_1 + \tilde{h}_2$, $\tilde{h}_{34} = \tilde{h}_3 + \tilde{h}_4$, $\tilde{h}_{1234} = \tilde{h}_{12} + \tilde{h}_{34}$.

Temporal hierarchies require  the mean and the variance  of the base forecasts.  The original algorithm \cite{athanasopoulos2017forecasting} provides only the reconciled point forecast, while the approach of  \cite{Corani2021}  yields also a   reconciled predictive  distribution.


\section*{Experiments}
\label{sec:casestudy}
The BigDEAL Challenge 2022 was divided in a qualifying match and a final match. 

\begin{figure}[ht!]
    \includegraphics[width=\textwidth]{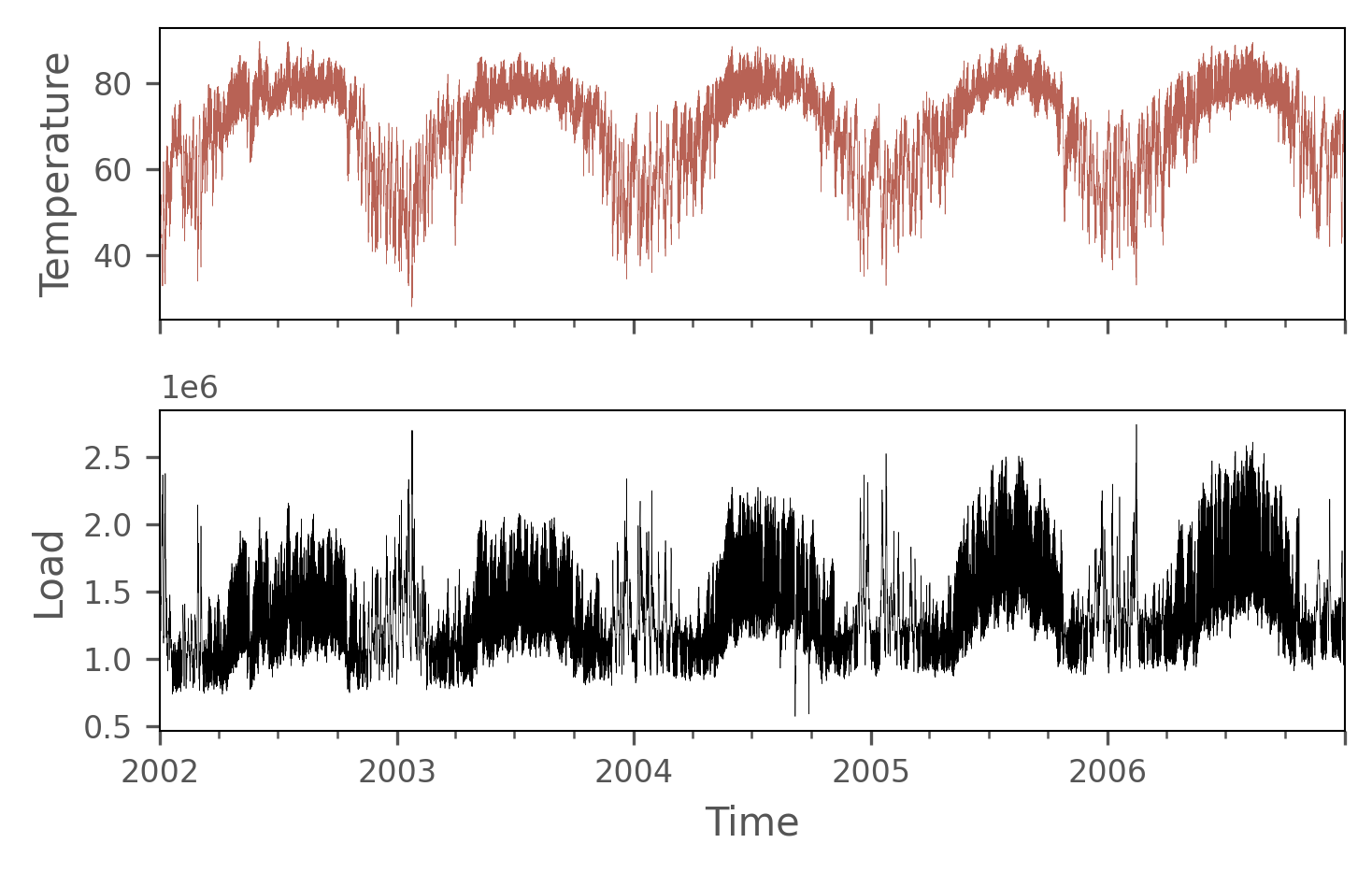}
    \caption{Load (MW) and temperature $\text{T}_{avg}$ (°F) of the qualifying match of the BigDEAL challenge. For readability, we show data aggregated over 12 hours.}
    \label{fig:qualidata}
\end{figure}

The \textit{qualifying match} provided hourly load and hourly temperature statistics (mean, median, min, max) of four weather stations for the period 2002-2006; see  Fig. \ref{fig:qualidata} for an  example.
It required to forecast year 2007
given the \textit{actual} temperatures. This is referred to as a \textit{ex-post} setting \cite{hyndman2009density}.

The \textit{final match} provided three years (2015-2017) of hourly load of three U.S. local distribution companies (LDC), and hourly temperatures  from six weather station. 
The \textit{forecasted} (\textit{ex-ante} setting \cite{hyndman2009density}) 1-day ahead temperatures for 2018 were released on a rolling basis, two months at a time. The forecasts for these periods were to be delivered, in a total of six consecutive rounds.

Both matches required forecasts at hourly scale for the 24h, the values of the peak for each day, and its time of occurrence (i.e. a discrete number between 1 and 24).


The organizers ran two  baseline methods.
The first is \textit{Tao's Vanilla Benchmark}  \cite{hong2014global}, namely  a multiple linear 
regression whose features are trend, calendar  effects, and only a single feature related to the temperature.
The second is the \textit{Recency Benchmark} \cite{wang2016electric}, which extends the previous model by including many features related to temperatures, among which lagged values and moving averages.


\subsection*{Performance measures}
\label{section:performance_measures}
The organizers evaluated the forecasts of each match with three different tracks.

In the qualifying match the hourly forecasts ($24\times365=8760$) were scored using the \textit{Mean Absolute Percentage Error} (MAPE):
\begin{equation}
    \text{MAPE} = \frac{1}{H}\sum_{t=T+1}^{T+H}{\frac{|y_t - \hat{y}_t|}{|y_t|} \times 100 },
    \label{eq:mape}
\end{equation}
where $y_t$ and $\hat{y}_t$ denote the actual and the forecasted value for time $t$. 
The second metrics was the \textit{Magnitude} (M); it is the MAPE between
the actual and forecasted daily peak values (i.e., it refers to 365 forecasts with one year horizon).
However, we recall that MAPE has been criticized  in the forecasting literature: it penalizes over-estimation errors more than under-estimation ones \cite{Armstrong1992} and moreover it is  numerically unstable when dealing with values close to 0. 

To score the prediction of peak hours the organizers used a third metric, called \textit{Timing} (T), which computes the \textit{Mean Absolute Error}. For example, if actual peak is at 6pm, and forecasted peak time is at 8pm, the error for that day is $|6-8|=2$.

The final match scored the forecasts using \textit{Magnitude} (M) and \textit{Timing} (T), plus an additional metric called \textit{Shape} (S). 
However, the definition of Timing was modified introducing a non-uniform cost for the error.
Let us denote by $\mathcal{T}_{d}$ and $\hat{\mathcal{T}}_{d}$ the actual and the forecasted peak hour for a day $d$. Timing was then defined as:
\begin{equation}
\begin{aligned}
    \text{T} = \frac{1}{\left|days\right|} \sum_{d\,\text{in}\, days} w(\mathcal{T}_{d}, \hat{\mathcal{T}}_{d}), \text{with} \\
    w(\mathcal{T}_{d}, \hat{\mathcal{T}}_{d}) =
    \begin{cases}
    |\mathcal{T}_{d} - \hat{\mathcal{T}}_{d}|,& \text{if } |\mathcal{T}_{d} - \hat{\mathcal{T}}_{d}| = 1,\\
    2|\mathcal{T}_{d} - \hat{\mathcal{T}}_{d}|,& \text{if } 2 \leq|\mathcal{T}_{d} - \hat{\mathcal{T}}_{d}| \leq 4,\\
    10,& \text{if } |\mathcal{T}_{d} - \hat{\mathcal{T}}_{d}| \geq 5
    \end{cases}
\end{aligned}
\label{eq:timing}
\end{equation}
\textit{Shape} (S) scored the shape of the forecast around the peak. To compute it, the 24h load forecasts of a day is normalized by the peak forecast of that day, and the same is done for the actual load. Then the sum of absolute errors during the 5-hour peak period (actual peak hour $\pm$ 2 hours) of every day is calculated.
We denote by $\bar{y}_{d}$ and $\bar{\hat{y}}_{d}$ the normalized actual and forecasted load for a day $d$; $\bar{y}_{d} = \frac{y_{d}}{\max{y_{d}}}$, $\bar{\hat{y}}_{d} = \frac{\hat{y}_{d}}{\max{\hat{y}_{d}}}$. Shape is defined as:
\begin{equation}
    \text{S} = \frac{1}{\left|days\right|} \sum_{d\,\text{in}\, days}
    \sum_{t\,\text{in}\,\{\mathcal{T}_{d}, \mathcal{T}_{d}\pm1, \mathcal{T}_{d}\pm2\}}
    |\bar{y}_{d}(t) - \bar{\hat{y}}_{d}(t)|
    \label{eq:shape}
\end{equation}

\paragraph{Scoring the Predictive Distribution} 
While the  competition only assessed the point forecasts, we also scored the distributional forecasts obtained from our probabilistic models.
In particular, we compared the probabilistic forecast of our GB model (based on  \cite{Marz2022}) with those obtained after the application of the temporal hierarchy.


We scored the predictive distributions of the model using the \textit{Continuous Ranked Probability Score} (CRPS) \cite{Gneiting2005}. Let us denote by $\mathbf{\hat{F}}$ the predictive cumulative distribution function and by $y$ the actual value:
\begin{equation}
    \text{CRPS}(\mathbf{\hat{F}}, y) = 
    \int_{-\infty}^{\infty} (\mathbf{\hat{F}}(x) - \mathbbm{1}(x\geq y))^2
    \,dx
    \label{eq:crps}
\end{equation}
With Gaussian $\mathbf{\hat{F}}$, the integral can be computed in closed form \cite{Gneiting2005}.

We then  scored the prediction intervals using the 
 \textit{Interval Score} (IS) \cite{gneiting2011quantiles}.
Let us denote by $1-\alpha$ the nominal coverage of the interval (assumed 0.9 in this paper), by $\mathbf{l}$ and $\mathbf{u}$ its lower and upper bound. 
We thus computed with the models, for each hour, a  $90\%$ prediction interval and the score:
\begin{equation}
    \text{IS}(\mathbf{l},\mathbf{u},y)= (\mathbf{u}-\mathbf{l}) + \frac{2}{\alpha}(\mathbf{l}-y)\mathbbm{1}(y < \mathbf{l}) + \frac{2}{\alpha} (y-\mathbf{u}) \mathbbm{1}(y > \mathbf{u})\,.
\end{equation}
We also report the proportion of cases in which the interval ($\mathbf{l}, \mathbf{u}$) contains $y$.

\paragraph{\textit{Skill score}}Let $m_{origin}$ and $m_{new}$ two comparable metrics. We denote the positive or negative percentage improvement by the Skill score defined as:
\begin{equation}
    \text{Skill}_{\%}(m_{origin},m_{new}) = \frac{m_{origin} - m_{new}}{(m_{origin} + m_{new})/2} \times 100
    \label{eq:improvement}
\end{equation}



\subsection*{Qualifying Match}
Below we present the building blocks of our implementation.

\paragraph{Cross-validation strategy}
Given the sequential nature of the data, time series cross-validation was used to evaluate the performance of each model, hyper-parameter tuning and feature selection. This method involves partitioning the data into an in-sample data set comprising earlier observations, and an out-of-sample data set containing the most recent observations. The model is trained on the in-sample part and evaluated on the out-of-sample part.
To obtain multiple evaluations, the out-of-sample set is \textit{iteratively} shifted forward by a time window, and the model is retrained on the updated in-sample set.
This procedure is repeated for all the out-of-sample sets, ensuring that the model is tested on unseen sequences of observations during evaluation, which results in a more realistic assessment. 

The size of the time window is typically chosen equal to the size of the test set on which the final prediction is to be made. Hence, for the qualifying phase, years 2004, 2005 and 2006 were used as out-of-sample folds.

\paragraph{Baseline model}
We started by modelling basic calendar features (\texttt{Year}, \texttt{Month}, \texttt{Week}, \texttt{Day}, \texttt{Weekday}, \texttt{Hour}) and temperatures at the current time (\texttt{$\text{T}_{avg}$}, \texttt{$\text{T}_{med}$}, \texttt{$\text{T}_{min}$}, \texttt{$\text{T}_{max}$}). Calendar variables were encoded with label encoding. We tested more sophisticated encoding  (target encoding, cyclical encoding with sine/cosine transformation and cyclical encoding with radial basis functions) but without notable improvement. 

\begin{figure}[ht!]
   \subfloat[\centering Forecasts]{{\includegraphics{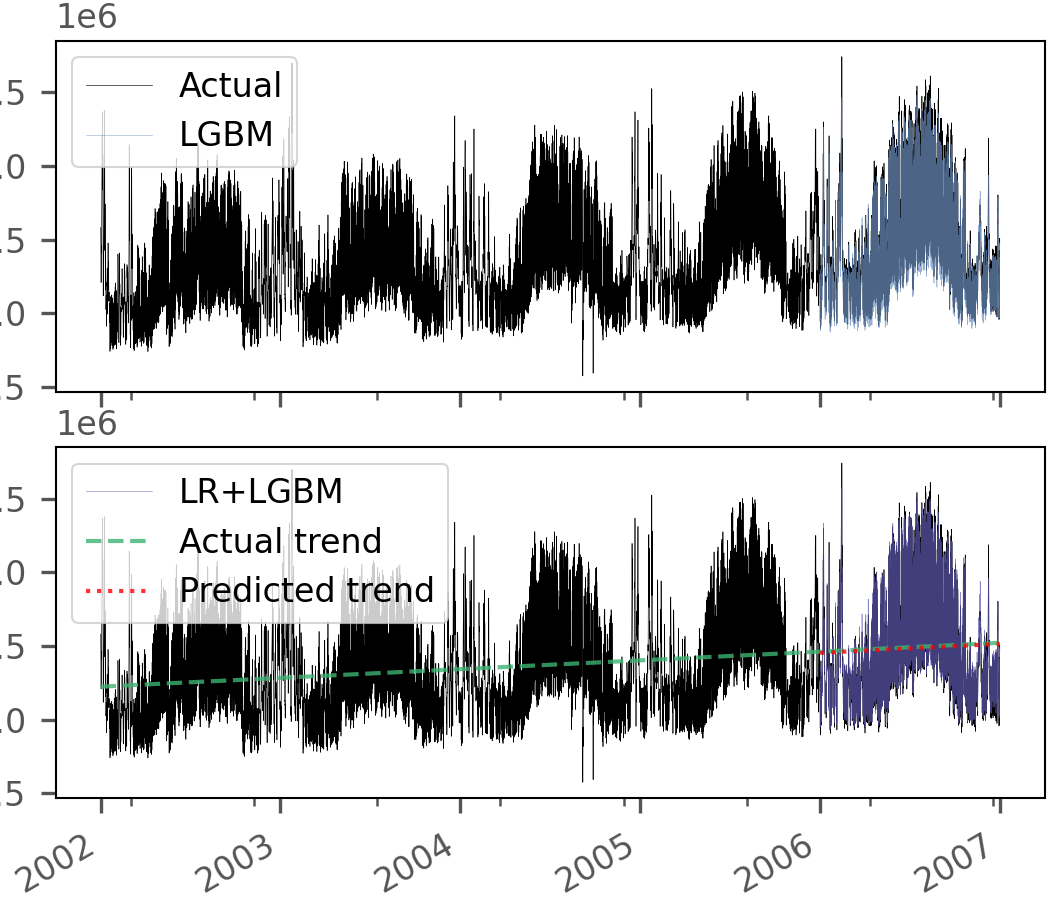} \label{fig:detrend_forecasts} }}%
   \hspace*{-0.6em}
   \subfloat[\centering Residuals]{{\includegraphics{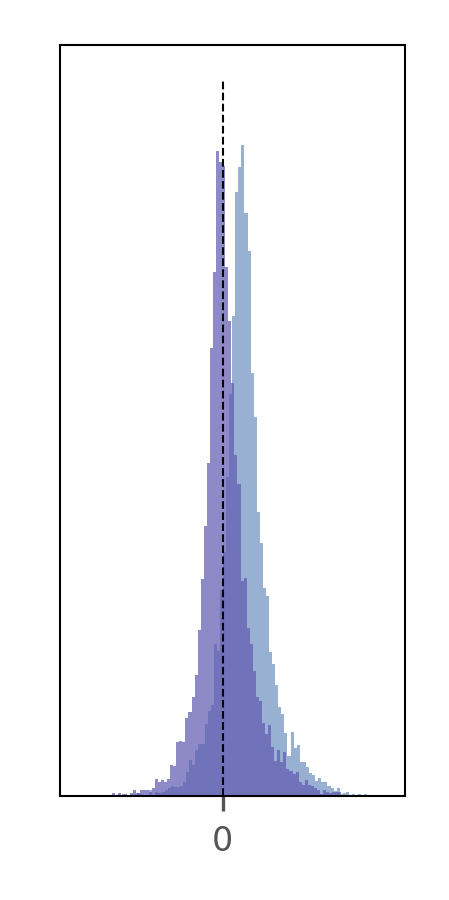} \label{fig:detrend_residuals} }}%
   \caption{The effect of  detrending. The LR model ($\beta_1 = 7.05$) corrects the underestimation.
   The plot of the residuals (on the right) shows how the distribution of the residuals is correctly centered in 0 after detrending.}%
   \label{fig:detrend}%
\end{figure}

We applied  a logarithmic transformation  to the target variable to stabilize its variance.
Moreover, since the target variable has a long-term increasing trend, we performed detrending.
During each round of cross-validation, we fitted a  Linear Regression (LR) model ($y_i = \beta_0 + \beta_1x_i$, where $x_i$ are progressive time indices with $i = 1,\ldots,T$) to the in-sample data.  We subtracted the  linear trend  from the training data before fitting the LightGBM model. At the prediction stage, we added the predicted linear trend  to the out-of-sample predictions, followed by an exponential transformation, to obtain the final forecast. The effect of detrending is shown in Fig. \ref{fig:detrend}. With  detrending the residuals of the models have mean 0 as expected; otherwise, they are severely biased. By detrending we  reduced
the MAPE (H) of the baseline model from 6.18 to 4.81.

\paragraph{Loss function} We trained the LightGBM models using the L2 loss function.

\paragraph{Outliers} To prevent the models to be biased by outliers, we assigned zero weight to time steps containing these values. We detected them from the left side of the target distribution, which we believe correspond to periods of severe weather conditions.

\paragraph{Feature engineering}
Feature engineering was performed \textit{incrementally} by adding related feature blocks one step at a time. This enabled a systematic evaluation of the impact of each block on the model's performance while ensuring that every new set of related features provided additional information. 
We found the following features  to be predictive for this competition:

\begin{itemize}
    \item[-] Additional calendar features: \texttt{Holiday}, \texttt{Holiday name}, \texttt{Weekend}, \texttt{Week of month}, \texttt{Season}, \texttt{Day of year}, \texttt{Days since last / until next holiday}.
    \item[-] Lagged hourly temperatures: for each temperature variable (\texttt{$\text{T}_{avg}$}, \texttt{$\text{T}_{med}$}, \texttt{$\text{T}_{min}$}, \texttt{$\text{T}_{max}$}) lagged hourly temperatures were incorporated into the model, ranging from a minimum lag of 1 hour to a maximum lag of 48 hours, for a total of 192 new features.
    \item[-] Temperature-based rolling statistics: for each temperature variable,
    and for 4 different values of window widths (3 hours, 1 day, 1 week, 1 month), 5 statistical functions (\textit{mean, max, min, median, std}) were computed, for a total of 80 new features.
    \item[-] Aggregated temperature statistics: for each temperature variable,
    for 2 different aggregation time period (Year-Month-Day, Month-Hour), 11 aggregation functions (\textit{mean, max, min, median}, and centered \textit{RMS, crest factor, peak value, impulse factor, margin factor, shape factor, peak to peak value}) coupled with the differences between the actual temperature values and the aggregated values were computed for a total of $88 \times 2=176$ new features.
    For clarity, we indicate as an example the daily maximum with the notation $\tilde{T}_{max}^{Year, Month, Day}(t)$, where Year-Month-Day indicates the aggregation to be read from left levels to right.
\end{itemize}

\paragraph{Feature selection}
To evaluate our feature selection strategy, we carried out multiple experiments. First, we assessed the model performance without any feature selection (experiment \textbf{a}). Then, we applied the feature selection strategy described in Sec. \nameref{sec:feature_selection} after completing all feature engineering, on the entire set of features added to the baseline model (experiment \textbf{b}). Finally, we performed \textit{step-by-step} feature selection whenever we added a new block of features to the model, i.e. after adding lagged variables, after adding rolling variables, and so on (experiment \textbf{c}).
To enhance reliability, cluster permutations were executed 100 times, and mean values and standard deviations of performance drops were calculated against all out-of-sample folds. We consider a cluster of features \textit{informative} if the importance values fall within three standard deviations of the mean above 0. 

Results are presented in Tab. \ref{tab:qualification_results}. Specifically, the columns for MAPE, Magnitude, and Timing present the results based on the respective competition metrics, whereas columns \textbf{a}, \textbf{b}, and \textbf{c} correspond to the 3 experimental strategies employed.

It is important to note that, unlike experiment \textbf{a}, where the results were obtained in a single training run, the results for experiment \textbf{b} and \textbf{c} were derived from three different training runs, each one maximizing the metric of interest.

\setlength{\tabcolsep}{6pt}
\begin{table}[ht]
\centering
\caption{Out-of-fold qualification results with feature selection methods.} 
\vspace{6pt}
\begin{tabular}{lccccccccc}
\multicolumn{1}{l}{} & \multicolumn{3}{l}{MAPE (H)} & \multicolumn{3}{l}{Magnitude} & \multicolumn{3}{l}{Timing} \\
\cmidrule(l){2-10}
\multicolumn{1}{l}{} & \textbf{a} & \textbf{b} & \textbf{c} & \textbf{a} & \textbf{b} & \textbf{c} & \textbf{a} & \textbf{b} & \textbf{c} \\ \midrule
Baseline  & 4.81 & - & -       & 4.43 & - & -       & 1.42  & - & - \\
Calendar  & 4.83 & - & 4.78    & 4.48 & - & 4.46    & 1.39  & - & 1.39 \\
Lags      & 3.33 & - & 3.24    & 3.29 & - & 3.20    & 0.94  & - & 0.92 \\
Roll lags & 3.28 & - & 3.16    & 3.22 & - & 3.20    & 1.06 & - & 0.94 \\
Agg stats & 3.24 & 3.16 & 3.09 & 3.21 & 3.10 & 3.08 & 0.91 & 0.95 & 0.91 \\ \bottomrule
\end{tabular}
\label{tab:qualification_results}
\end{table}

\begin{figure}[ht!]
    \centering
    \subfloat[\centering Dendrogram]{{\includegraphics{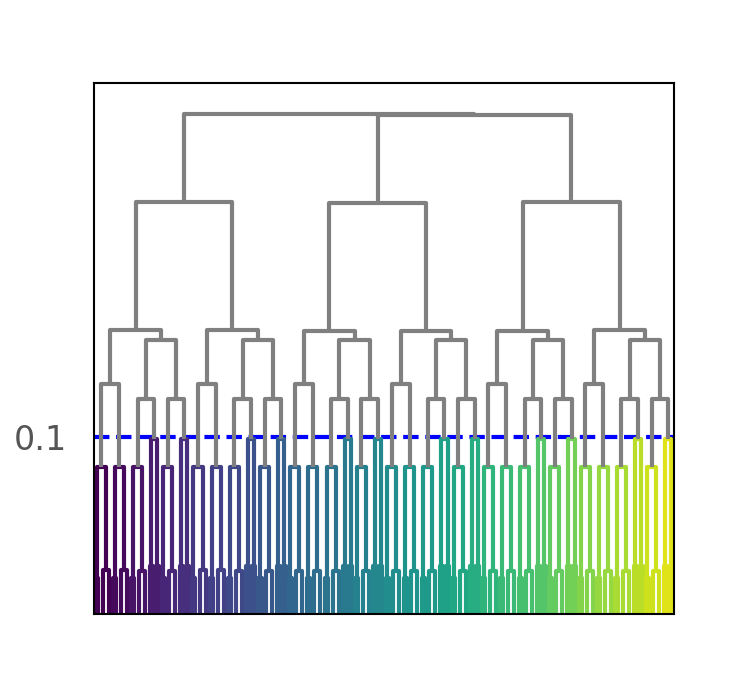} \label{fig:dendrogram} }}%
    \hspace*{-0.6em}
    \subfloat[\centering Correlation matrix]
    {{\includegraphics{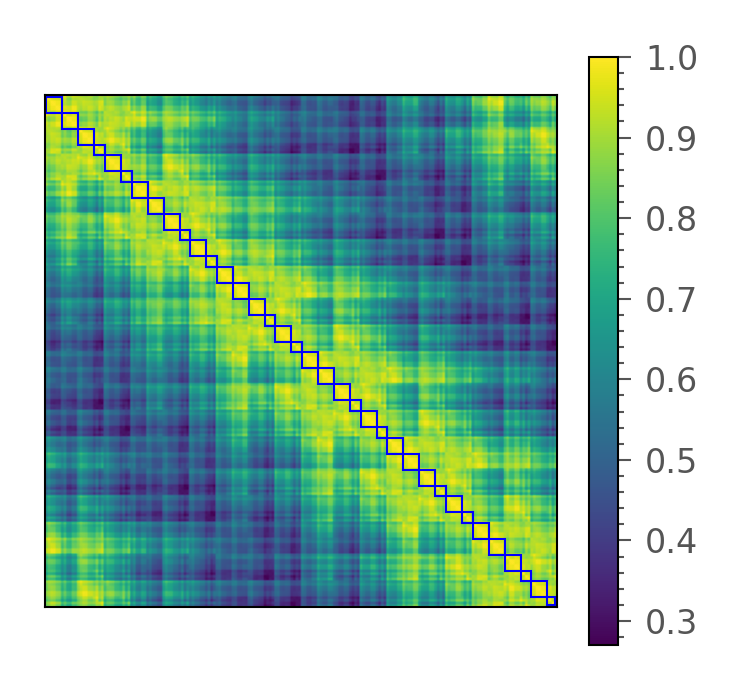}} \label{fig:correlation_matrix} }%
    
    \subfloat[\centering CPFIs]{{\includegraphics{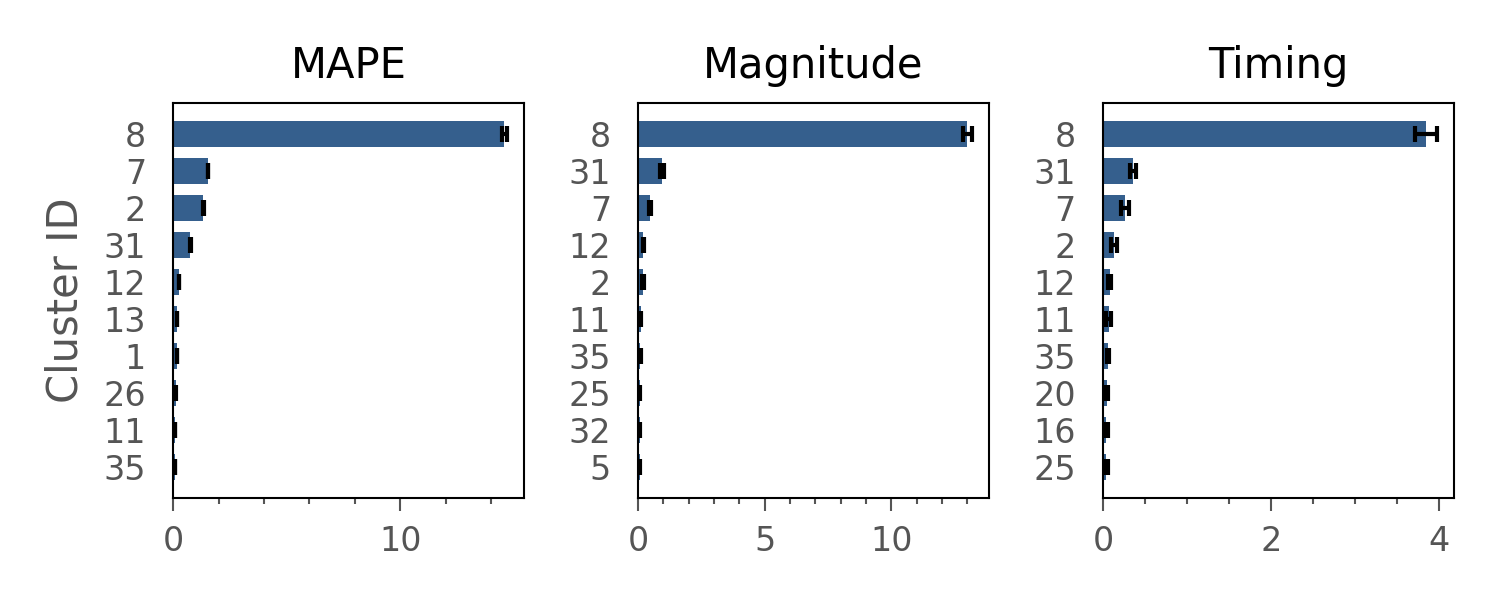}} \label{fig:cpis} }%
    \caption{Hierarchical  clustering   (0.1 threshold) (a) and Spearman's correlation matrix (b). The blue squares highlight the 36 clusters. In (c) Clustered Permutation Feature Importance (CPFI) values are reported for each track. See Tab. \ref{tab:cluster_rank} for the Top-5 feature sets.}%
    \label{fig:clustering}%
\end{figure}

For illustration purposes, in Fig. \ref{fig:clustering}, we present the feature selection results obtained after incorporating lagged hourly temperatures into the model.
Fig. \ref{fig:dendrogram} presents the dendrogram obtained from hierarchical clustering computed on the Spearman correlation matrix shown in Fig. \ref{fig:correlation_matrix}. By selecting a threshold value of 0.1, we identified 36 clusters. 
The cluster rankings that maximize, respectively, the performance of MAPE, Magnitude, and Timing are visible in Fig. \ref{fig:cpis}. For all the metrics, cluster 8 proved to be the most significant, followed by clusters 31, 7, 2 and 12. This suggests that most informative lags are at t-$\{1,2,3,4,5,6\}$, t-$\{11,12\}$ and t-$\{25,26\}$. Tab. \ref{tab:cluster_rank} shows the clusters associated feature set. 

\setlength{\tabcolsep}{6pt}
\begin{table}[ht]
\centering
    \caption{Clustered Permutation Feature Importance: Top-5 clusters of lagged temperatures that maximize performance indicators.}
    \vspace{6pt}
    \begin{tabular}{cl}
         Cluster ID & Feature Set  \\ \midrule
         8 & \texttt{$\text{T}_{avg, med, min}(t-1)$, $\text{T}_{avg, med, min}(t-2)$} \vspace{0.1cm}\\
         31 & \texttt{$\text{T}_{avg, med, min}(t-11)$, $\text{T}_{avg, med, min}(t-12)$} \vspace{0.1cm}\\
         7 & \texttt{$\text{T}_{avg, med, min}(t-3)$, $\text{T}_{avg}(t-4)$} \vspace{0.1cm}\\
         2 & \texttt{$\text{T}_{avg, med, min}(t-5)$, $\text{T}_{avg, med}(t-6)$} \vspace{0.1cm}\\
         12 & \texttt{$\text{T}_{max}(t-1)$, $\text{T}_{max}(t-2)$, $\text{T}_{max}(t-25)$, $\text{T}_{max}(t-26)$} \\
         \bottomrule
    \end{tabular}
    \label{tab:cluster_rank}
\end{table}

\paragraph{Hyper-parameter optimization} We used Optuna  \cite{akiba2019optuna} to tune the learning parameters of LightGBM. It implements time-budget optimization which was useful given the short deadlines of the competition. 
Another strength of  Optuna is that it allows to optimize the parameters with respect to the metric of the competition.

\paragraph{Results} 
Our team was named ``swissknife'';
as reported in Tab.~\ref{tab:qualileaderboard}, 
we ranked $8^{th}$ on the hourly forecast (H), $3^{rd}$ on the Magnitude (M), $3^{rd}$ on the Timing (T).

\setlength{\tabcolsep}{1pt}
\begin{table}[!ht]
\tiny
\centering
\caption{Leaderboard of Qualifying Match \cite{BigDEALChallQuali}.} \vspace{6pt}
\begin{tabular}{c@{\hskip -0.1cm}c@{\hskip -0.1cm}c@{\hskip -0.1cm}c@{\hskip -0.1cm}c@{\hskip -0.1cm}c}
\textbf{Team} & \textbf{Rank H.} & \textbf{Team} & \textbf{Rank M.} & \textbf{Team} & \textbf{Rank T.} \\ \midrule
X-Mines     & 1     & Amperon       & 1     & RandomForecast    & 1 \\ 
Amperon     & 2     & Team SGEM KIT & 2     & Amperon           & 2 \\
Yike Li     & 3     & \cellcolor{yellow}swissknife    & 3     & \cellcolor{yellow}swissknife        & 3 \\
peaky-finders & 4   & peaky-finders & 4     & freshlobster      & 4 \\
KIT-IAI     & 5     & KIT-IAI       & 5     & peaky-finders     & 5 \\
Overfitters & 6     & EnergyHACker  & 6     & \cellcolor{orange}\textit{Recency Benchmark} & \\
BelindaTrotta & 7   & BelindaTrotta & 7     & X-Mines           & 6 \\
\cellcolor{yellow}swissknife  & 8     & Overfitters   & 8     & BrisDF            & 7 \\
\cellcolor{orange}\textit{Recency Benchmark} & & VinayakSharma & 9 & BelindaTrotta     & 8 \\
RandomForecast  & 9 & SheenJavan      & 10    & KIT-IAI           & 9 \\
Team SGEM KIT   & 10 & \ldots        &    & SheenJavan        & 10 \\
\ldots & & \cellcolor{orange}\textit{Recency Benchmark} & 13 & \ldots & \\
\cellcolor{sienna}\textit{Tao's Vanilla Benchmark} & 27 & \cellcolor{sienna}\textit{Tao's Vanilla Benchmark} & 25 & \cellcolor{sienna}\textit{Tao's Vanilla Benchmark} & 30 \\
\end{tabular}
\label{tab:qualileaderboard}
\end{table}

\subsection*{Final Match}
For the final match, we followed the same pipeline tuned in the qualification phase, with the exception of target transformation, which was not required as the target variable was already stationary. 
Additionally, three LDC loads were to be forecasted (\texttt{LDC1, LDC2, LDC3}) instead of one, and the temperature variables come directly from six weather stations (\texttt{T1, T2, T3, T4, T5, T6}), without aggregate statistics, and moreover without geographical references.

To further enhance performance, we incorporated several techniques, including DART, probabilistic LightGBM and temporal hierarchies.

\begin{figure}[ht!]
    \centering
    \includegraphics{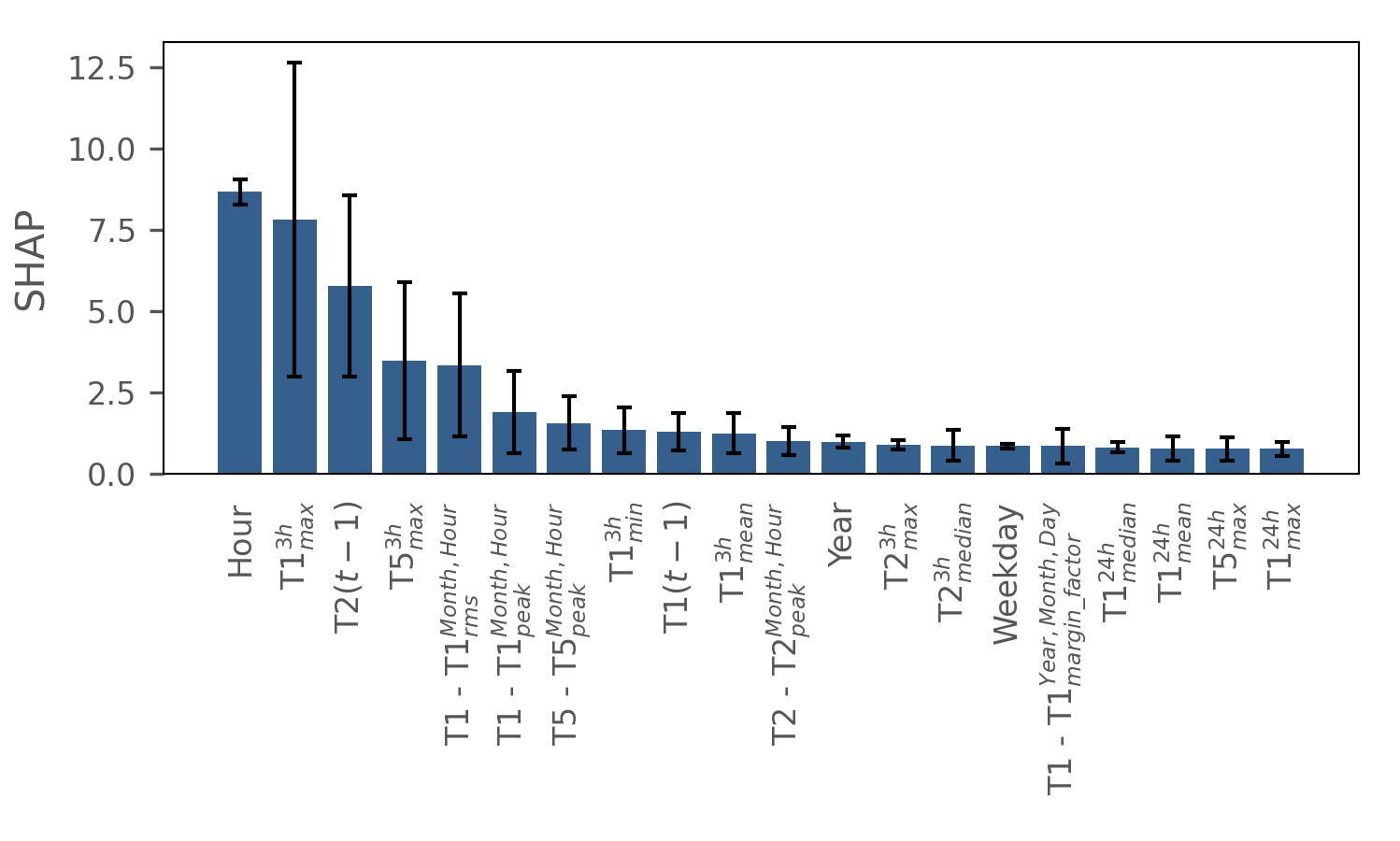}
    \caption{Out-of-fold Top-20 Features Importance obtained after the last incremental step of feature engineering (\textit{aggregated features}) and feature selection, for \texttt{LDC1} at Round 1. On the \textit{y}-axis we reported SHAP (SHapley Additive exPlanations) \cite{lundberg2017unified} values of the LightGBM model.}
    \label{fig:FIFinalRound1LDC1}
\end{figure}

\paragraph{Feature selection}
The most important lagged temperatures were found at time t-$\{1,2,3,4,5\}$, and t-$\{10,11,12\}$ and the  most important rolling lag temperatures were found with $w=\{3\,hours,\,1\,day\}$.  Fig. \ref{fig:FIFinalRound1LDC1}  shows
that within that six weather station, temperatures \{\texttt{T1, T2, T5}\} better explain \texttt{LDC1}. Hence, our framework   nicely handles dataset with multiple weather station.

\paragraph{Regularization}
The use of DART booster reduce the overfitting that affects LightGBM with standard booster by means of \textit{Dropout}. 
However, training becomes slower since it requires more boosting iterations.
Additionally, early stopping cannot be used because the algorithm updates previous learned tree during the training. 
Despite these issues, with DART 
we reduced the prediction error. 
Also we find out that the loss decreases steadily and converges asymptotically, even with a large number of boosting iterations.
We tested DART on the qualification data only when it was finished. With 30'000 iterations, the MAPE (H) went from 3.24 to 2.83, and Magnitude from 3.21 to 3.09. Therefore, we included DART in the final match model.

\paragraph{Temporal hierarchies} 
We build a temporal hierarchy by summing the time series of the hourly load and related temperatures at the following scales: \textit{2-hours, 4-hours, 6-hours, 12-hours}. 
We train an independent probabilistic LightGBM-LSS \cite{Marz2022} model at each time scale, with DART booster for $30'000$ boosting steps. LightGBM-LSS minimizes the Negative Log-Likelihood loss function.
Gaussian distributional forecasts were obtained at each temporal scale for the same forecasting horizon $H$. 
The reconciliation step made base forecasts coherent. 

\begin{figure}[ht!]
    \includegraphics{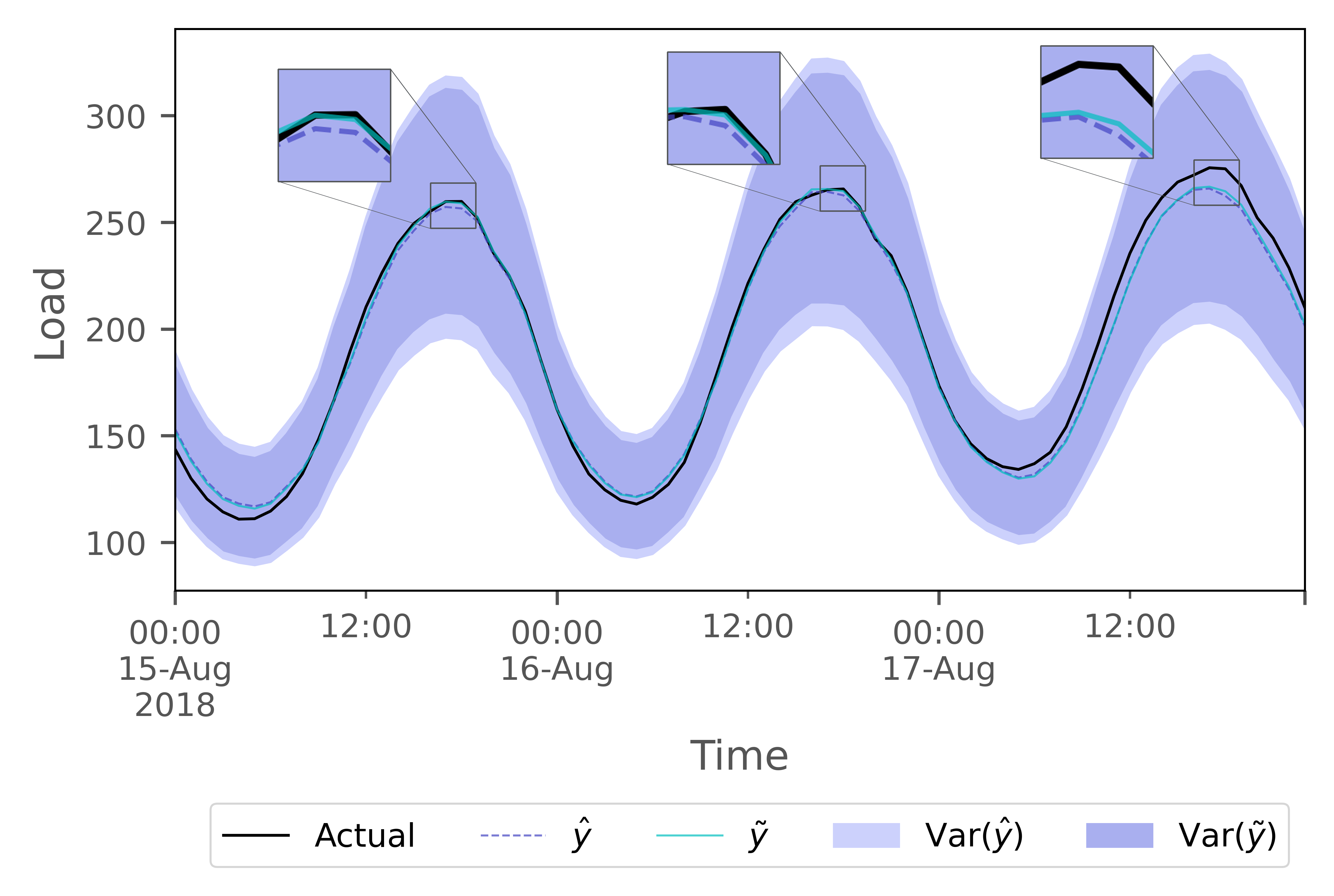}
    \caption{Comparison of probabilistic forecasts, before and after the application of the temporal hierarchy. 
    The temporal hierarchy slightly improves the point forecasts.  It also shortens the prediction intervals without compromising their reliability.  
    The sample refers to  three days  (15-17 Aug 2018) for  
    \texttt{LDC1}. 
    }
    \label{fig:probfore}
\end{figure}

In Tab. \ref{tab:cs_reconctable_load} (load profile) and Tab. \ref{tab:cs_reconctable_peak} (peak) we compare base and reconciled forecasts, using skill scores ($S_{\%}$); in Fig. \ref{fig:probfore} we show some forecasts.
Temporal hierarchies improve only  slightly the point forecasts, but more importantly  the predictive distribution, with a skill score of about 5\% on CRPS and of about 10\% on IS.

We also tested 1-day aggregation without further improvement for the bottom time series. As can be seen from previous feature importance analyses, past values close to the conditioning time are the most important variables for prediction. We come to the explanation that too much aggregation (empirically greater then 1-day) makes these variables vanishing, and temporal hierarchy no longer effective. Instead, small hierarchies also improve peaks, as shown in Tab. \ref{tab:cs_reconctable_peak} and Fig. \ref{fig:probfore}.

Given the availability, the metrics we present for the final match refer to actual competition values of Round 1-5 (Jan-Oct 2018), instead of out-of-fold sets.

\setlength{\tabcolsep}{4.5pt}
\begin{table}[ht!]
\centering
\scriptsize
\caption{Reconciliation metrics for the \textit{load profiles}; base ($\hat{y}$) and reconciled ($\tilde{y}$) forecasts, with skill scores ($S_{\%}$). Temporal hierarchy for forecasting using hourly (bottom level), 2-hourly, 4-hourly, 6-hourly and 12-hourly aggregations.} \vspace{6pt}
\begin{tabular}{crrB|rrB|rrB|rr}
\multicolumn{1}{l}{} & \multicolumn{3}{l}{ MAPE} & \multicolumn{3}{l}{ CRPS} & \multicolumn{3}{l}{ $\text{IS}_{90\%}$} & \multicolumn{2}{l}{ $\text{IC}_{90\%}\,(\%)$} \\ 
\cmidrule(l){2-12} 
\multicolumn{1}{l}{} & \multicolumn{1}{c}{$\hat{y}$} & \multicolumn{1}{c}{$\tilde{y}$} & \multicolumn{1}{c}{\cellcolor{aliceblue}$\mathbf{S}_{\%}$} & \multicolumn{1}{c}{$\hat{y}$} & \multicolumn{1}{c}{$\tilde{y}$} & \multicolumn{1}{c}{\cellcolor{aliceblue}$\mathbf{S}_{\%}$} & \multicolumn{1}{c}{$\hat{y}$} & \multicolumn{1}{c}{$\tilde{y}$} & \multicolumn{1}{c}{\cellcolor{aliceblue}$\mathbf{S}_{\%}$} & \multicolumn{1}{c}{$\hat{y}$} & \multicolumn{1}{c}{$\tilde{y}$} \\ \midrule
LDC1    & 4.87 & \textbf{4.84} & 0.75 & 6.35 & \textbf{6.03} & 5.16 & 61.62 & \textbf{55.01} & 11.34 & 99.24 & 98.81  \\
LDC2    & 5.02 & \textbf{4.99} & 0.52 & 10.92 & \textbf{10.44} & 4.49 & 101.35 & \textbf{90.39} & 11.43 & 99.07 & 98.49  \\
LDC3    & 4.51 & \textbf{4.5} & 0.05 & 45.99 & \textbf{43.84} & 4.78 & 446.49 & \textbf{398.37} & 11.39 & 98.85 & 98.14 \\ \bottomrule
\end{tabular}
\label{tab:cs_reconctable_load}
\end{table}

\setlength{\tabcolsep}{4pt}
\begin{table}[ht!]
\centering
\scriptsize
\caption{Reconciliation metrics for the \textit{peaks}; base ($\hat{y}$) and reconciled ($\tilde{y}$) forecasts, with skill scores ($S_{\%}$). Temporal hierarchy for forecasting using hourly (bottom level), 2-hourly, 4-hourly, 6-hourly and 12-hourly aggregations.} \vspace{6pt}
\begin{tabular}{crrB|rrB|rrB|rrB}
\multicolumn{1}{l}{} & \multicolumn{3}{l}{ Magnitude} & \multicolumn{3}{l}{ Timing} & \multicolumn{3}{l}{ Shape} & \multicolumn{3}{l}{ $\text{CRPS}_{\text{peak}}$} \\ \cmidrule(l){2-13} 
\multicolumn{1}{l}{} & \multicolumn{1}{c}{$\hat{y}$} & \multicolumn{1}{c}{$\tilde{y}$} & \multicolumn{1}{c}{\cellcolor{aliceblue}$\mathbf{S}_{\%}$} & \multicolumn{1}{c}{$\hat{y}$} & \multicolumn{1}{c}{$\tilde{y}$} & \multicolumn{1}{c}{\cellcolor{aliceblue}$\mathbf{S}_{\%}$} & \multicolumn{1}{c}{$\hat{y}$} & \multicolumn{1}{c}{$\tilde{y}$} & \multicolumn{1}{c}{\cellcolor{aliceblue}$\mathbf{S}_{\%}$} & \multicolumn{1}{c}{$\hat{y}$} & \multicolumn{1}{c}{$\tilde{y}$} & \multicolumn{1}{c}{\cellcolor{aliceblue}$\mathbf{S}_{\%}$} \\ \midrule
LDC1    & 4.97  & \textbf{4.90} & 1.34 & 1.22 & \textbf{1.13} & 7.93 & 0.088 & \textbf{0.086} & 2.16 & 8.33  & \textbf{7.89}  & 5.46 \\
LDC2    & 5.51  & \textbf{5.48} & 0.52 & 1.26 & \textbf{1.23} & 1.87 & 0.102 & \textbf{0.101} & 1.11 & 15.73  & \textbf{15.13}  & 3.85 \\
LDC3    & 4.83  & \textbf{4.79} & 0.95 & 1.19 & \textbf{1.09} & 8.80 & 0.079 & \textbf{0.078} & 1.56 & 60.83 & \textbf{57.97} & 4.82 \\ \bottomrule
\end{tabular}
\label{tab:cs_reconctable_peak}
\end{table}

\paragraph{Results} We placed $6^{th}$ (M), $6^{th}$ (T), $7^{th}$ (S), see Tab. \ref{tab:finalleaderboard}.

\setlength{\tabcolsep}{1pt}
\begin{table}[ht!]
\centering
\tiny
\caption{Leaderboard of Final Match \cite{BigDEALChallFinal}.} \vspace{6pt}
\begin{tabular}{c@{\hskip -0.1cm}c@{\hskip -0.1cm}c@{\hskip -0.1cm}c@{\hskip -0.1cm}c@{\hskip -0.1cm}c}
\textbf{Team} & \textbf{Rank M.} & \textbf{Team} & \textbf{Rank T.} & \textbf{Team} & \textbf{Rank S.} \\ \midrule
Amperon & 1                   & KIT-IAI & 1                   & KIT-IAI   & 1 \\
Overfitters & 2               & Amperon & 2                   & Amperon   & 2 \\
peaky-finders & 3             & BelindaTrotta & 3             & Overfitters & 3 \\
Team SGEM KIT & 4             & Overfitters & 4               & X-mines     & 4 \\
KIT-IAI & 5                   & X-mines & 5                   & SheenJavan  & 5 \\
\cellcolor{yellow}swissknife & 6                & \cellcolor{yellow}swissknife & 6                & Rajnish Deo & 6 \\
\cellcolor{orange}\textit{Recency Benchmark} & 7  & peaky-finders & 7             & \cellcolor{yellow}swissknife  & 7 \\
Energy HACker & 8             & Rajnish Deo & 8               & \cellcolor{orange}\textit{Recency Benchmark} & 8 \\
Rajnish Deo & 9                & Team SGEM KIT & 9            & RandomForecast & 8.5 \\
X-mines & 10                   & SheenJavan & 10            & Yike Li & 8.5 \\
\ldots   & &                       \ldots & &                   peaky-finders & 10 \\
\cellcolor{sienna}\textit{Tao's Vanilla Benchmark} & 17.5 &    \cellcolor{orange}\textit{Recency Benchmark} & 14 & \ldots & \\
&                           & \cellcolor{sienna}\textit{Tao's Vanilla Benchmark} & 18 & \cellcolor{sienna}\textit{Tao's Vanilla Benchmark} & 16 \\ \\
\end{tabular}

\begin{tabular}{c@{\hskip -0.1cm}c@{\hskip -0.1cm}}
\textbf{Team} & \textbf{Final Rank} \\ \midrule
Amperon         & 1 \\
KIT-IAI         & 2 \\
Overfitters     & 3 \\
peaky-finders   & 4 \\
X-mines         & 5 \\
\cellcolor{yellow}swissknife      & 6 \\
Rajnish Deo     & 7 \\
Team SGEM KIT   & 9 \\
\cellcolor{orange}\textit{Recency Benchmark} & 10 \\
\ldots & \\
\cellcolor{sienna}\textit{Tao's Vanilla Benchmark} & 14
\end{tabular}
\label{tab:finalleaderboard}    
\end{table}

\section*{Conclusion}
\label{sec:conclusion}
In this work we shared our experience which allowed us to obtain 
excellent results in an international energy forecasting competition.
Our major contribution  is the definition of large set of 
explanatory variables, some borrowed from the literature  of signal processing, and a novel strategy for feature selection we called Clustering Permutation Features Importance (CPFI). 
The art of feature engineering was made necessary by the choice of 
using regression models based on GB.
In the recent past, especially the implementations of XGboost and LightGBM 
have proven to be valid and somehow superior to competing Neural Networks.
Then we consolidated the potential of LightGBM as a regressor for 
load forecasting, in a traditional tabular application.
We point out the improvement that DART booster allowed us to achieve over the traditional Gradient Boosting (GB) of trees.
The second major contribution of this paper is the implementation of
the recent probabilistic extension of LightGBM which returns the moments of a
predictive distribution, instead of the point forecast solely. This 
has a great
impact because the decision maker can also rely on the uncertainty inherent 
in the forecast. The popularity of these models is still limited, 
even in energy forecasting.
With distributional forecasts we were able to apply temporal hierarchies
and further improve the results. The competition did not assessed prediction uncertainty, but our probabilistic approach also proved to be calibrated in the intervals.




\subsection*{Availability of data and materials} 
The competition dataset is not publicly available at the date of the submission. The organizers plan to release it eventually.


\subsection*{Competing interests}
The authors declare that they have no known competing financial interests or personal relationships that could have appeared to influence the work reported in this paper.

\bibliographystyle{abbrv} 
\bibliography{EI23}      

\end{document}